\documentclass{eceasst}
\usepackage{pifont} 
\usepackage{stmaryrd} 
\usepackage{amssymb}
\usepackage{amsmath}
\usepackage[latin1]{inputenc}  
\volume{X}{2012} 
\volumetitle{
Proceedings of the\\
12th International Workshop on\\
Automated Verification of Critical Systems\\
(AVoCS 2012)}
\volumeshort{
Proc.\ AVoCS 2012}
\guesteds{
Gerald Lüttgen, Stephan Merz}

\title{Towards Focus on Time} 
\short{Towards  Focus on Time} 
\author{Maria Spichkova\autref{1}} 
\institute{\autlabel{1} \email{spichkov@in.tum.de}, \url{http://www4.in.tum.de/~spichkov/}\\
Institut für Informatik, TU München,\\ 
Boltzmannstr. 3, 85748 Garching, Germany} 
\abstract{On specifying and verifying timing system properties in a formal way.} 

\begin{document}
\maketitle
\newcommand{\ntsplit}[2]{\ensuremath{#1\curlywedgedownarrow_{#2}}}
\newcommand{\ntjoin}[2]{\ensuremath{#1\curlyveedownarrow_{#2}}}
\newcommand{\Focus}{\textsc{Focus} }


Specifying real-time systems, arguing about timing properties is hardly avoidable. 
The notion of time takes center stage for this kind of systems: 
abstracting from the time we may loose the core properties of a system we represent, e.g.\ in many cases the causality property can be loosed. 
In particular, the timed domain is the most important one for representation 
of distributed systems with real-time requirements. 

On the other hand, in many cases arguing about time makes the specification more readable for an average engineer:
specification of a real-time system in an untimed frame may be in some cases 
shorter or more elegant from mathematical point of view, but 
 to understand such specifications and to argue about their properties is in many cases 
much more difficult in comparison to the corresponding specifications 
in a timed frame, where some properties like causality could be represented explicitly. 
Moreover, abstraction from timing aspects can easily lead to specification 
mistakes because of difficulties to find a correct abstraction.

One of the most well-established models for the specification and verification of real-time system design is 
representation of a system using timed automata,  introduced by Alur and Dill~\cite{Alur94atheory} --   them extend the classical finite automata by a finite set of real valued clocks used to measure the time elapsed between events to constrain the runs of the automation. 
The  model from Alur and Dill is well-developed, however, some its assumptions on the system could be too idealistic: 
timed automata assumes perfect continuity of clocks and, what could be even more critical if we deal with embedded systems, instantaneous reaction times. Another problem of this model is that it does not prevent Zeno runs~\cite{ZenoRuns}. To solve this, the idea of robust model checking was introduced by  Puri~\cite{Puri} and also revised in other approaches. 

We suggest another solution: to argue not about a single action or messages but about time intervals containing a finite number of actions, i.e.\ 
to argue not about single messages in an input stream, but about a sequence 
of messages that are present in this stream at some time interval. This sequence can be in general either empty or contain a single message or a number of messages.  
In this approach a discrete model of time is used, because $(i)$ this simplifies the specification and, even more, the verification of system properties, and $(ii)$ any timed transition system can be discretized without loss of generality. %
Thus, an input for our timed automata or, more generally, a timed component, is an infinite sequence of finite time intervals. 

However a discrete model of time is used here, ones can choose
any time granularity defining the concrete meaning of a time interval according to the  system requirements, and, moreover, 
it is possible to switch from one time granularity to another: the operator \ntsplit{s}{n} refines the time granularity  splitting every time interval of the stream $s$ into $n$ time intervals (there is a number of strategies to apply for the split action), where the operator \ntjoin{s}{n} makes the time granularity more coarse -- 
it joins $n$ time intervals of the stream $s$ into a single time interval.
We can use any granularity defining the concrete meaning of a time interval according to the  system requirements, but one of the main problems with the timed automata -- Zeno runs -- still be excluded. 

To make specification more readable and to simplify the verification of the corresponding system properties, we introduced, in addition to the representation as state machines (timed automata), also  a special kind of tables and a number of other operators for the formal specification of real-time systems (as well as auxiliary lemmata over them): from untimed causal simulation and timed merge to the operators for the analyzing of crypto-based distributed software and their composition. 
To avoid misinterpretations, we call our kind of timed automata by \emph{timed state transition diagrams} (TSTDs). Specifying system behavior by TSTD we can use three specification styles: diagram, table and a  textual style. 

The TSTD-approach can be applied within any specification/verification language, nevertheless we prefer to present it using \Focus~\cite{focus}, 
a framework for formal specification and development of interactive systems, to take advantage of its features, inter alia the well-developed theory of composition and  
extensions based on human factor analysis within formal methods~\cite{hffm_spichkova}. %
Using this approach, one can verify properties of a system  in a formal way according to the methodology ``\Focus~on Isabelle''\cite{spichkova}, 
by translating the \Focus~specifications to the semiautomatic theorem prover Isabelle/HOL~\cite{npw}, an interactive semi-automatic theorem prover,  and using the Isabelle tool to make the proofs.

The feasibility of the proposed approach   
has been demonstrated by two  industrial case studies from automotive area, motivated   
by DENSO Corporation and Robert Bosch GmbH. 

\bibliographystyle{eceasst}

\begin{thebibliography}{NPW02}

\bibitem[AD94]{Alur94atheory}
R.~Alur, D.~L. Dill.
A Theory of Timed Automata.
\emph{Theoretical Computer Science} 126:183--235, 1994.

\bibitem[BS01]{focus}
M.~Broy, K.~St{\o}len.
\emph{Specification and Development of Interactive Systems: Focus on Streams,
  Interfaces, and Refinement}.
Springer, 2001.

\bibitem[GB07]{ZenoRuns}
R.~G\'{o}mez, H.~Bowman.
Efficient detection of Zeno runs in timed automata.
In \emph{Proceedings of the 5th international conference on Formal modeling and
  analysis of timed systems}.
FORMATS'07, pp.~195--210.
Springer-Verlag, 2007.

\bibitem[NPW02]{npw}
T.~Nipkow, L.~C. Paulson, M.~Wenzel.
\emph{{Isabelle/HOL -- A Proof Assistant for Higher-Order Logic}}.
LNCS~2283.
Springer, 2002.

\bibitem[Pur00]{Puri}
A.~Puri.
Dynamical Properties of Timed Automata.
\emph{Discrete Event Dynamic Systems} 10(1-2):87--113, 2000.

\bibitem[Spi07]{spichkova}
M.~Spichkova.
\emph{{Specification and Seamless Verification of Embedded Real-Time Systems:
  FOCUS on Isabelle}}.
PhD thesis, {TU M{\"u}nchen}, 2007.

\bibitem[Spi12]{hffm_spichkova}
M.~Spichkova.
{Human Factors of Formal Methods}.
In \emph{{In IADIS Interfaces and Human Computer Interaction 2012 (IHCI
  2012)}}.
2012.

\end{thebibliography}

\end{document}